\documentclass[11pt,a4paper]{article}
\usepackage{jheppub}
\pdfoutput=1
\usepackage[T1]{fontenc}
\usepackage[utf8]{inputenc}
\usepackage{lmodern}
\usepackage{amsmath,amsfonts,amssymb}
\usepackage{hyperref}
\usepackage[all]{hypcap}
\usepackage{graphicx}
\usepackage{fixme}
\fxsetup{draft,theme=color}
\fxsetface{margin}{\scriptsize}
\usepackage{amsthm}

\usepackage{paralist}
\usepackage{subfigure}
\usepackage[textsize=scriptsize,backgroundcolor=red!70,linecolor=red]{todonotes}

\newcommand{\ie}{\textit{i.e.}}
\newcommand{\eg}{\textit{e.g.}}
\newcommand{\figref}[1]{Fig.\ \ref{fig:#1}}
\newcommand{\figlabel}[1]{\label{fig:#1}}

\title{Forward charm-production models and\\ prompt neutrinos at IceCube}

\author[a]{Atri Bhattacharya,}
\author[a]{J.~R.~Cudell}
\affiliation[a]{Space sciences and Technologies for Astrophysics Research (STAR) Institute, University of Li{\`e}ge, Li{\`e}ge, Belgium}

\emailAdd{A.Bhattacharya@uliege.be}
\emailAdd{JR.Cudell@uliege.be}
\abstract{  We investigate the prompt neutrino background at IceCube, as
  determined from forward charm.
  We consider the role of intrinsic charm and of a recombination model
  and show that the contribution of these mechanisms is at most a factor
  two.
}


\begin{document}

\maketitle

\section*{Introduction}
This century is witnessing the birth of a new astronomy making use of all the fundamental interactions 
to explore hidden processes in the Universe: the multi-messenger approach which enjoyed its 
first successes with the observation of GW170817/GRB170817A \cite{GBM:2017lvd}, and with
the recent IceCube breakthrough  \cite{IceCube:2018cha} which identified the blazar TXS 0506+056 as the source of IceCube-170922A.
This method will facilitate easier characterisation of sources, as well as
provide a new tool to discover unknown objects \cite{Abbott:2016blz}.
It may also open a window to particle physics discoveries
\cite{Bhattacharya:2014yha,Bhattacharya:2016tma}.
Important players in this multi-messenger program are the neutrino telescopes 
IceCube and ANTARES/KM3NET, which use large volumes of ice/water in the Earth.
Their main drawback is that the Earth comes with an atmosphere, which
produces significant background from the collisions of cosmic
rays on the air. These collisions, via the weak decay of light mesons, produce a large neutrino
background. However, at high energy, these light mesons interact with air molecules
before having the chance to decay. Hence, the neutrino background is increasingly suppressed with 
higher energies and higher time dilation.
However, there is one exception to this---mesons containing heavy quarks
will quickly decay and produce neutrinos.
Heavier quarks are produced through more energetic gluons, which are scarce
compared to low-energy ones.
This means that $ b\bar{b} $ and $t\bar t$ production is suppressed compared to $ c\bar{c} $.
Consequently, the main background at high energies comes from decays of charmed mesons.

The highest-energy neutrino background comes from forward production, and
this implies that the charm mass is the only available scale for perturbative QCD.
Inclusive QCD calculations of quark production start to make sense at the
charm mass scale but require at least a next-to-leading order evaluation
\cite{Nason:1987xz,Nason:1989zy,Bhattacharya:2015jpa,Bhattacharya:2016jce}.
However, we are interested in the semi-inclusive calculation as the charm quark is
singled out at the high momentum, and this implies we are sensitive to
non-perturbative effects which cannot be factored out into fragmentation 
or structure functions.
This makes the highest-energy charm background model dependent.
Recent perturbative calculations of charmed-meson production
are already consistent with observations from the latest collider experiments,
including ALICE, ATLAS, and CMS at low rapidities\footnote{The LHCb data, at central
rapidities, are well reproduced at $\sqrt{s}=$5 TeV and 7 TeV. At 13 TeV, it  may undershoot the prediction
although one should note that the data have been revised twice and the preprint five times \cite{Aaij:2015bpa}. 
If this result is confirmed, it will only strengthen our conclusion.}
This only leaves a small window of rapidities where the
diffractive cross-section can significantly contribute, \textit{viz.} at
very high rapidities.

There are currently no data corresponding to $ D^{\pm,0} $ production at
rapidities beyond those accessible to LHCb at $ \sqrt{s}=13 $ TeV;
as a consequence the constraints on theoretical parameters governing
the forward production of these mesons are rather weak.
At these very high rapidities, two classes of models are usually considered
to describe the production of heavy mesons:
\begin{inparaenum}[\itshape a\upshape)]
\item models where the charm can intrinsically
carry a large momentum that can then be inherited by the D mesons \cite{Brodsky:1980pb,Barger:1981rx,Vogt:1994zf},
and
\item those
that assume that the D is boosted because its light quark is a spectator valence quark from one of the
initial nucleons \cite{Braaten:2001bf,Braaten:2001uu,Braaten:2002yt,Norrbin:1998bw}.
\end{inparaenum}
The former is straightforwardly implemented via modified structure functions, whereas
the latter require dedicated computations involving new
fragmentation mechanisms.
Our goal in this work is to estimate the maximum prompt atmospheric neutrino flux
that may be seen at IceCube by allowing the parameters in these models to go as high
as possible whilst still maintaining consistency with data at central
rapidities.

This paper is organised as follows: in Section I, 
we discuss the predictions of the leading-twist NLO charm production calculation for very high rapidities, and
the corresponding uncertainties involved.
In Section II, we consider and constrain the possible
modification to the charm momentum from intrinsic charm \cite{Brodsky:1980pb}. In Section III, we consider the recombination process of Braaten, Jia 
and Mehen \cite{Braaten:2001bf,Braaten:2001uu,Braaten:2002yt}.
Finally, in Section IV, we estimate the corresponding prompt neutrino fluxes,
and show that this component of the background remains small in all cases.
\section{High-rapidity D-meson from the high-{\sl x}$_F$ NLO contribution}
As previously noted, the perturbative treatment of Bhattacharya-Enberg-Reno-Sarcevic-Stasto (BERSS) \cite{Bhattacharya:2015jpa,Bhattacharya:2016jce}
 has been tuned to the
experimental results for D-meson production cross-sections across a wide range of energies.
At the low-energy end of the spectrum, this includes data from
fixed-target, proton and pion beam experiments at the CERN SPS
at a few hundred GeV, and the 920 GeV fixed-target experiment HERA-B;
further up the energy ladder are the somewhat discordant results from STAR and PHENIX
fixed-target experiments, both involving a beam of energy 200 TeV; and  
finally, constraints at the highest energies come from
D-meson production data at the LHC at $ \sqrt s = 7 \text{ and } 13 $ TeV.
The total number of experimental points is thirteen.

The tunable parameters in the theory are the charm mass and the factorisation
and renormalisation scales.
By fitting the theoretical cross-sections, computed perturbatively to the next-to-leading
order, to observed data at these widely disparate energies, one is able to
obtain the best-fit and limiting values of these scales consistent with
observations. The overall fit has a $ \chi^2 $/d.o.f. of 1.31.
Allowing the factorisation and renormalisation scales to run proportionally
to the transverse mass of the charm quark $ m_T = \sqrt{m_c^2 + p_T^2} $, where
$ p_T $ is its transverse momentum, the BERSS analysis finds best fit values
at $ (M_F/m_T, \mu_R/m_T) = (2.1, 1.6) $ for the choice of fixed $ m_c = 1.27 $ GeV.
The upper and lower limits, determined with reference to $ 1\sigma $ errors on
experimental data, are shown in \cite{Bhattacharya:2015jpa}.
\section{Intrinsic Charm}
\begin{figure}
  \centering
  \subfigure{\includegraphics[width=0.25\textwidth]{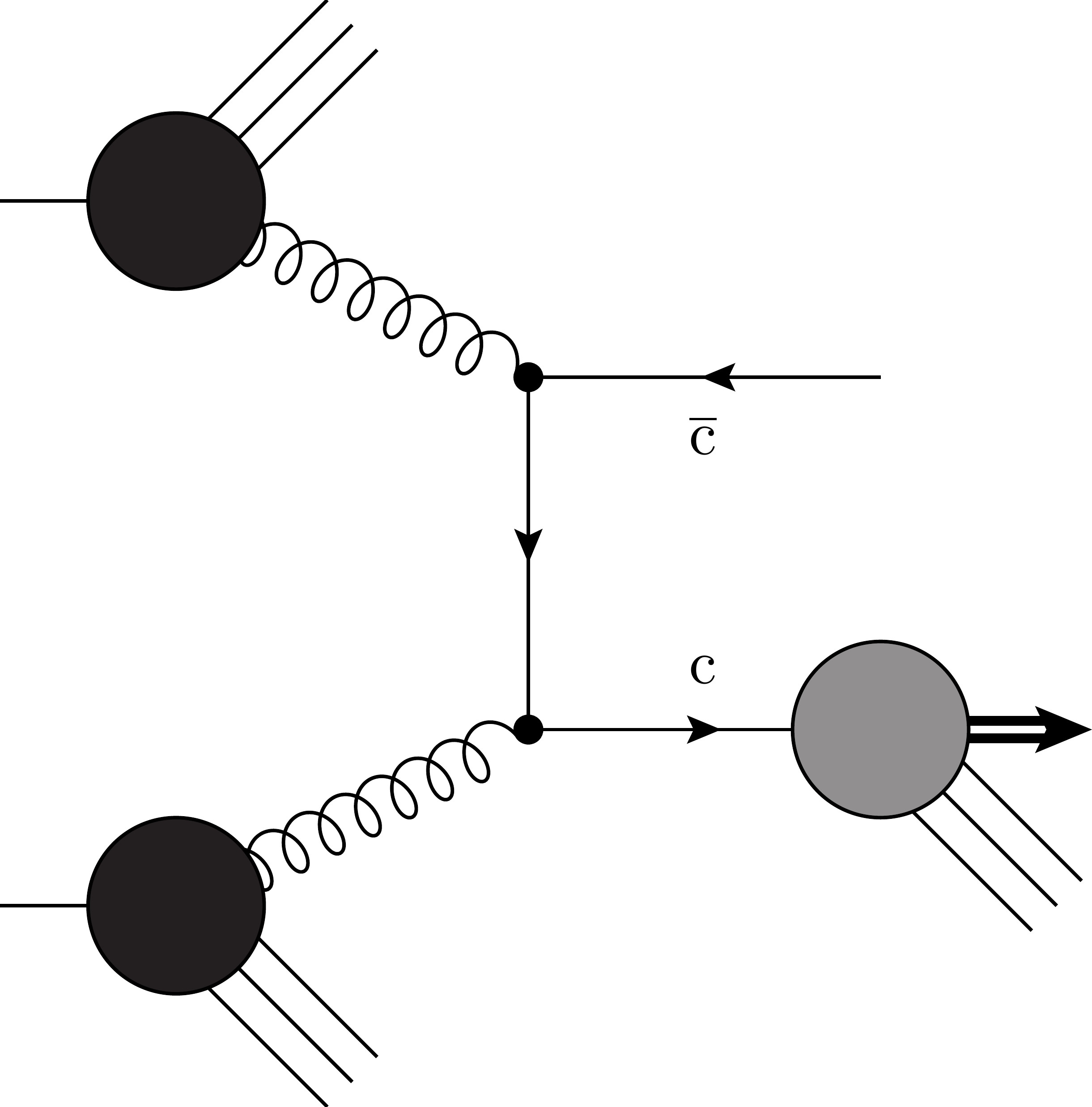}}\qquad\qquad
  \subfigure{\includegraphics[width=0.25\textwidth]{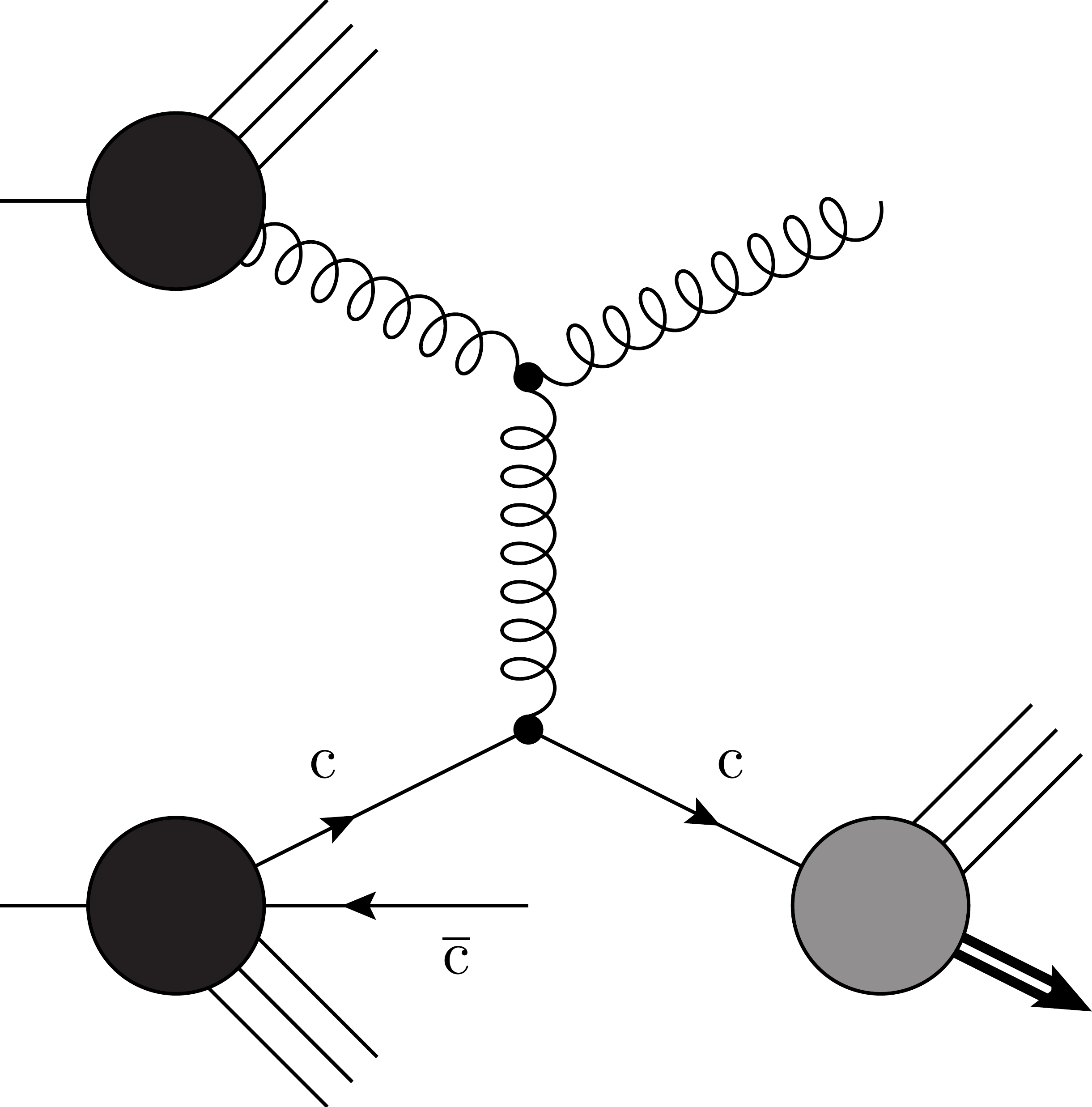}}\qquad\qquad
  \subfigure{\includegraphics[width=0.25\textwidth]{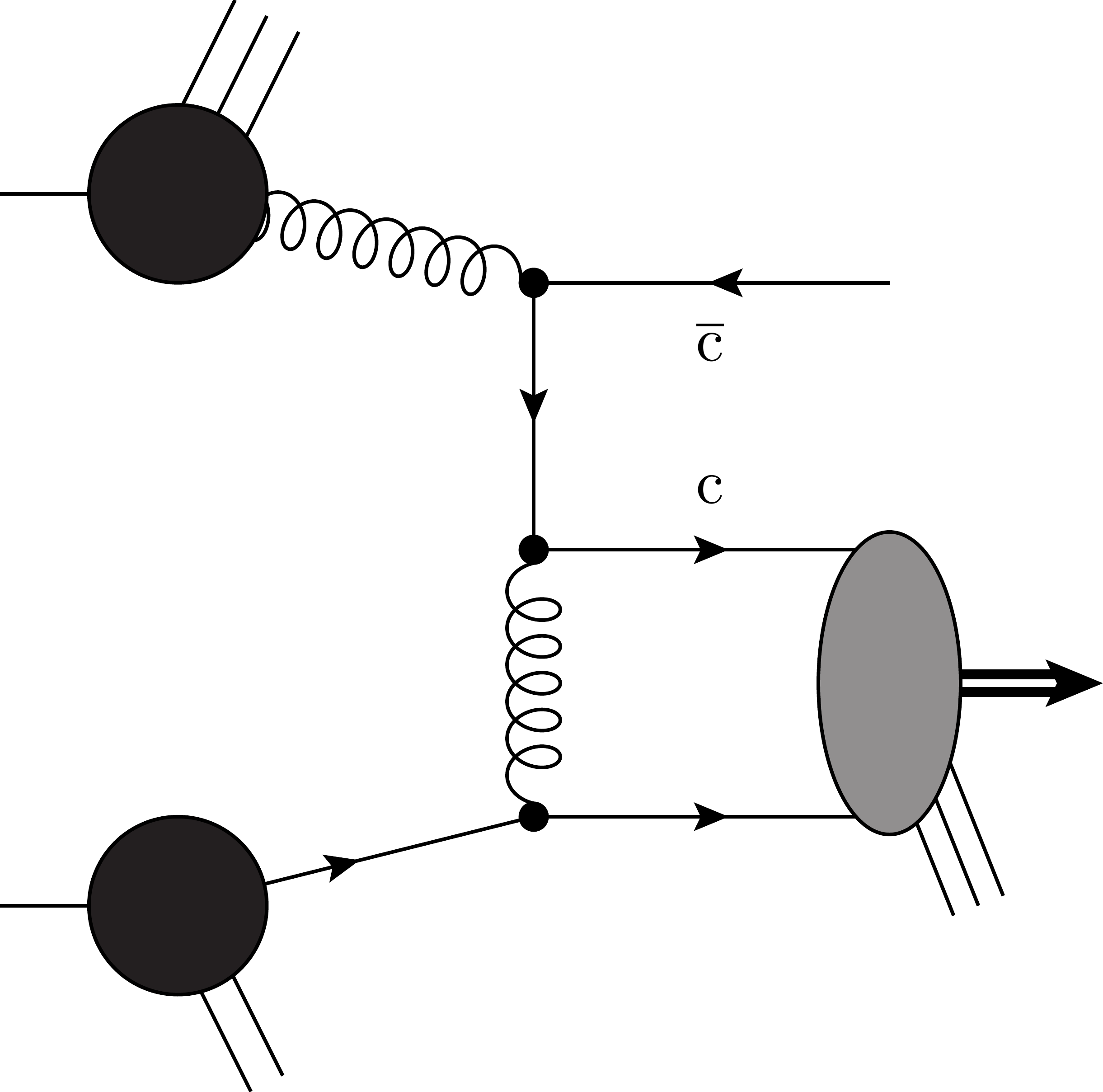}}
  \caption{\label{fig:interactions}Different processes leading to forward-charmed-meson production: (from left to right) perturbative
           gluon fusion in the vein of BERSS followed by fragmentation, intrinsic-charm-initiated $ D $-meson production followed by fragmentation, and
          charm-pair production leading
           to $ D $-mesons via recombination.}
\end{figure}
\begin{figure}[htb]
  \centering
  \includegraphics[width=0.85\textwidth]{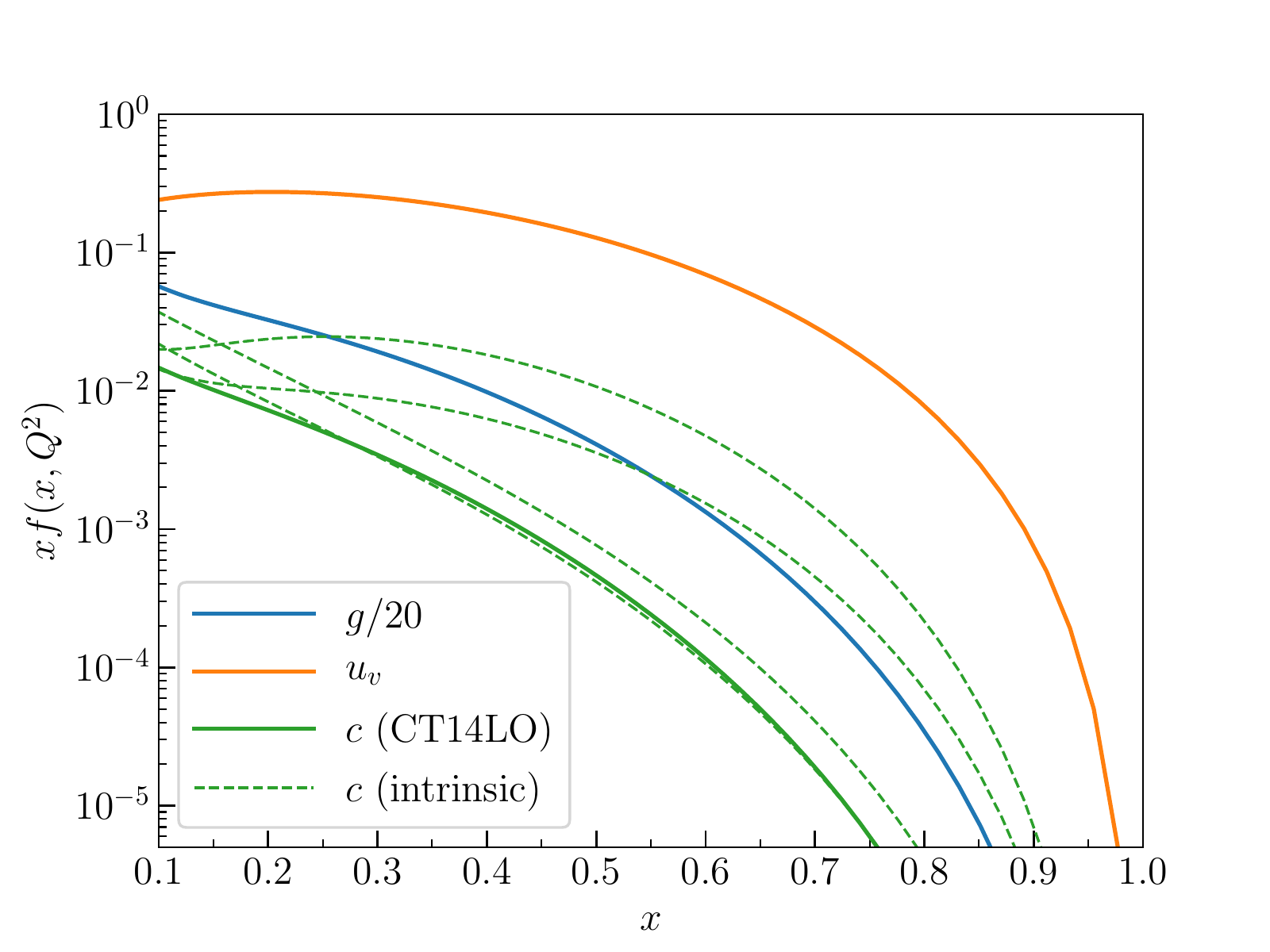}
  \caption{\label{fig:pdf-berss}Leading-order CT14 parton distributions
           for gluons, valence up quarks and charm quarks at $ Q^2 = 4m_c^2 $.
           Also shown (dashed green curves) are the four intrinsic charm parametrisations as part of the {\tt CT14LO-IC} set.
           }
\end{figure}

The simplest way to boost charm is to recognise that sometimes the proton
has a fluctuation that produces a $ c\bar{c} $ pair, as shown in Fig.~\ref{fig:interactions}.
This was recognised a long time ago \cite{Brodsky:1980pb}
and lead to  the introduction of intrinsic charm.
It is the part of the structure functions that comes
from the non-perturbative large-distance gluon field
and it cannot be described via DGLAP evolution.
Modern structure functions include this contribution as an initial parametrisation in some
of their sets.
Here we will consider the latest parametrisation due to the CT collaboration \cite{Hou:2017khm}.
They have looked at the intrinsic charm content in the
proton at the leading (LO), next-to-leading (NLO) and next-to-next-to-leading orders (NNLO). 
The structure functions are fit to the open-charm production
data in the central rapidity region from recent collider experiments.
The CT14 sets propose four distinct parametrisations of the
intrinsic charm distribution function (shown in Fig.~\ref{fig:pdf-berss}), and we choose the `BHPS2' set, which has the highest charm content
at large $ x $.

As the contribution of intrinsic charm will turn out to be small, it is sufficient to look at the leading-order calculation, for which we use the expressions of 
\cite{Vogt:1994zf}.
\begin{figure}[htb]
\centering
\includegraphics[width=0.85\textwidth]{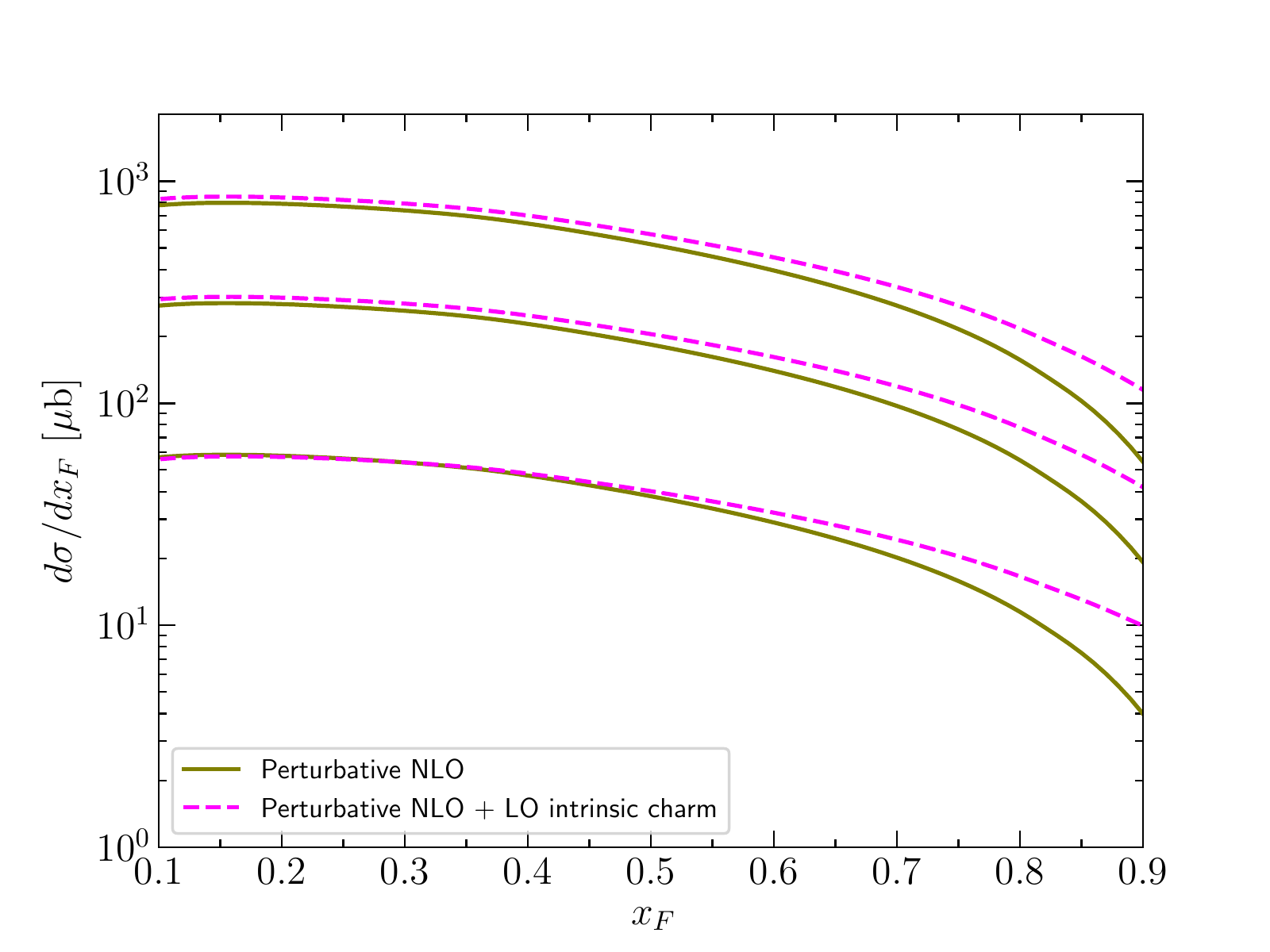}
\caption{\figlabel{sigIC}: Total D-meson production cross section $ d\sigma/dx_F $ as a function of the final-state meson $ x_F $
without and with intrinsic charm at $ \sqrt{s} =  10^{2}, 10^{3}, \text{ and } 10^{4}  $ GeV from bottom to top.}
 \end{figure}
There is a large uncertainty attached to the value of the quark mass, and we choose here a small charm mass, $ m_c = 1.27 $~GeV, that corresponds to 
a high
neutrino flux, as in \cite{Bhattacharya:2014yha,Bhattacharya:2016tma}.
We also use leading-order structure functions, and keep the Kramer-Kniehl fragmentation functions \cite{Kniehl:2006mw}.
The results are shown in Fig.\ 3, for several representative values of $ \sqrt{s} $,
where we add the intrinsic charm contribution to the NLO gluon-fusion cross section.
We see that intrinsic charm makes 
a difference only at very high $x_F>0.9$ 
where it can change the differential cross section by
a factor of two of larger.
However, this is the region of $x_F$ in which the differential cross section is suppressed, and we shall see that the effect on the neutrino 
flux is small.\footnote{This is even before taking into account the kinematic suppression of the charm pair production
cross-section from intrinsic charm as discussed in Ref.~\cite{Blumlein:2015qcn}.
This would appreciably reduce its contribution to the neutrino flux.}
Our prediction for the prompt neutrino flux when including intrinsic charm contribution
is consistent with results from other recent works \cite{Laha:2016dri,Giannini:2018utr,Bai:2018xum}.
It is significantly below the intrinsic charm contribution to the flux discussed in \cite{Halzen:2016thi},
which uses $ \Lambda^{0,\pm} $ production data from fixed-target experiments to deduce constraints on $ D^{0,\pm} $
production assuming extreme values of fractional momentum transferred to the final
$ \bar{c}c $ pair.
As such, the flux derived therein must be seen as a weak upper limit
to the intrinsic charm contribution and our results are well within this limit.
Note also that the NLO calculation is infrared finite and includes the mechanism considered in \cite{Halzen:2016pwl}
without the need for an infrared cut-off.

\section{\label{sec:braaten-form}The BJM recombination formalism}
In the Braaten-Jia-Mehen (BJM) formalism for heavy-quark  recombination with an active light quark, 
the $ D $-production cross section is expressed to leading order as product of two factors:
\begin{equation}\label{eq:braatenccbar}
  d\hat{\sigma}\left[D\right] = d\hat{\sigma}\left[\bar{q}g \to \left[\bar{q}c(n)\right] + \bar{c}\right]
                                \times \rho\left[\left[\bar{q}c(n)\right] \to D\right].
\end{equation}
The first factor is the usual perturbative term. 
Whereas the standard calculation (BERSS) calculates this factor to the next-to-leading order and then
multiplies it by fragmentation functions to make $ D $-mesons,
an extra contribution is considered here, in which the light quarks do not come from the vacuum but rather
from the proton.
This contribution is parametrised by the non-perturbative
numerical factor $ \rho\left[\left[\bar{q}c(n)\right] \to D\right] $, which is calculated from an
effective Lagrangian.
The quantity $ n $ is representative of the colour and spin quantum numbers of the $ \bar{q}c $ pair.
This formalism allows for coloured bound states or spin-flips.
For example, the production of a colour singlet, 
spin-conserving
$D^{+} $ meson is given by
the product of
\begin{align}\label{BJM}
  \frac{d\hat{\sigma}}{dt}[\bar{d}c({^1S_0^{(1)}})] &=
{2 \pi^2 \alpha_s^3 \over 243} {m_c^2 \over S^3}\left[-{64\,U \over S}
+{m_c^2\over T}\left(79 -{112 \,U\over T} -{64 \,U^2 \over T^2}\right)\right.\\
& \left.\qquad \qquad \qquad -{16\,m_c^4 \,S \over U\, T^2}
\left(1- {8\,U\over T} \right)\right] , \nonumber
\end{align}
and
$ \rho^\text{sm}_{1} = \rho\left[\bar{d} c\left({}^{1}\!S^{(1)}_0\right) \to D^{+}\right] $,
where, ``sm'' indicates a spin-matched pair, \ie\ $ c $ and $ \bar{d} $ have matching spins,
and $ S,\, T\text{, and } U $ are modified Mandelstam variables that can be expressed
in terms of the final state transverse momentum $ p_\perp $ and rapidity $ \Delta y $ as
\begin{align*}
  S &= 2(p_\perp^2+m_c^2)(1+{\rm cosh} \Delta y )\, , \\
  T &= -(p_\perp^2+m_c^2)(1+e^{-\Delta y})\, , \\
  U &= -(p_\perp^2+m_c^2)(1+e^{\Delta y}) \, .
\end{align*}
Expressions for other cross-sections are given in \cite{Braaten:2001bf,Braaten:2001uu}.
Note that if one includes fragmentations  $\left[\bar{d}c\left({}^{1}\!S^{(1)}_0\right) \to D^{+}X\right] $ then
$\rho$ becomes a free positive parameter, that can be adjusted to the data.
BJM found that one only needs the colour-singlet spin-conserving term to reproduce
the high-$ x_F $ charge asymetries in $D^+$/$D^-$ production observed in $\pi^- p$ by the E791 experiment, with a value $\rho_1=0.06$. They also fitted 
photoproduction data, which leads to $\rho_1\approx 0.15$. 
More recently, these expressions have been used 
\cite{Lai:2014xya}
to reproduce the  $ D$ asymmetry observed at LHCb. There it was found, at somewhat lower values of $x_F$, that coulour-singlet spin-non-flip terms 
were not enough to reproduce the asymmetry. As we concentrate on the high-$x_F$ data, we shall adopt the BJM result, and consider the interval
[0.06, 0.15] for $\rho$.

\begin{figure}[htb]
\centering
\includegraphics[width=0.95\textwidth]{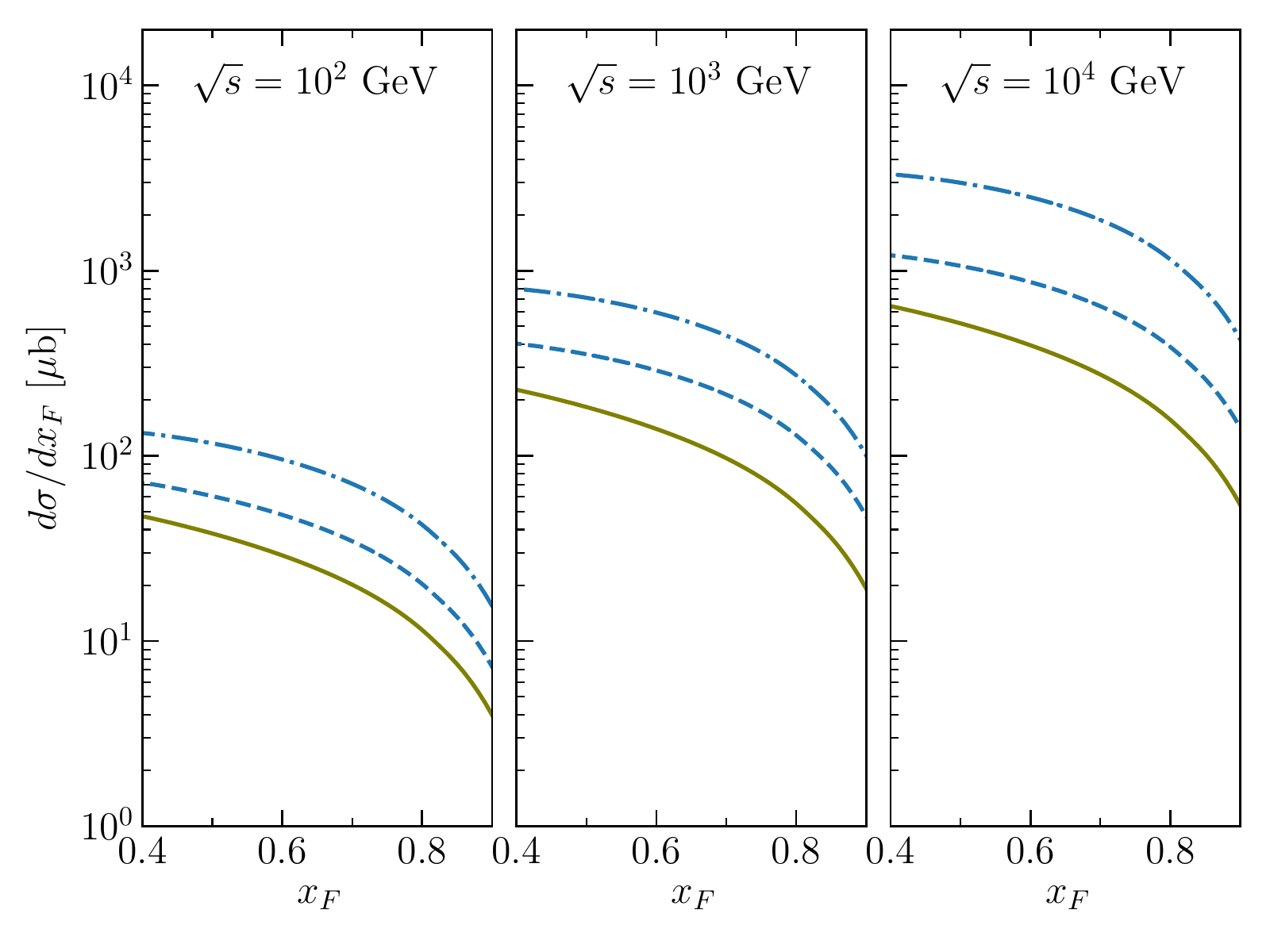}
\caption{\figlabel{sigBJM}: Total $D$-meson production cross section $ d\sigma/dx_F $ as a function of the final-state meson $ x_F $.
The NLO perturbative calculation is shown as the green, solid curve and the differential cross-section when including the BJM model
with $ \rho = 0.06\ (0.15) $ is shown as dashed (dot-dashed), blue
curve for $ \sqrt{s} = 10^{2}, 10^{3}, \text{ and } 10^{4} $ GeV.}
 \end{figure}

We see in Fig. 4 that the inclusion of the recombination enhances
the forward $ D^{\pm,0} $ production by up to a factor of 2 for the
higher value of $ \rho = 0.15 $ at $ x_F \approx 0.9 $.
Note that the BJM model is valid at high $x_F$ only and that it may get substantial corrections 
at low and medium values of $x_F$, hence the beginning of the curve of Fig. 4 is only indicative.
\section{\label{ssec:braaten-prflux}Prompt neutrino fluxes from the best-fit cross-section}
\begin{figure}[htb]
  \centering
  \includegraphics[width=0.7\textwidth]{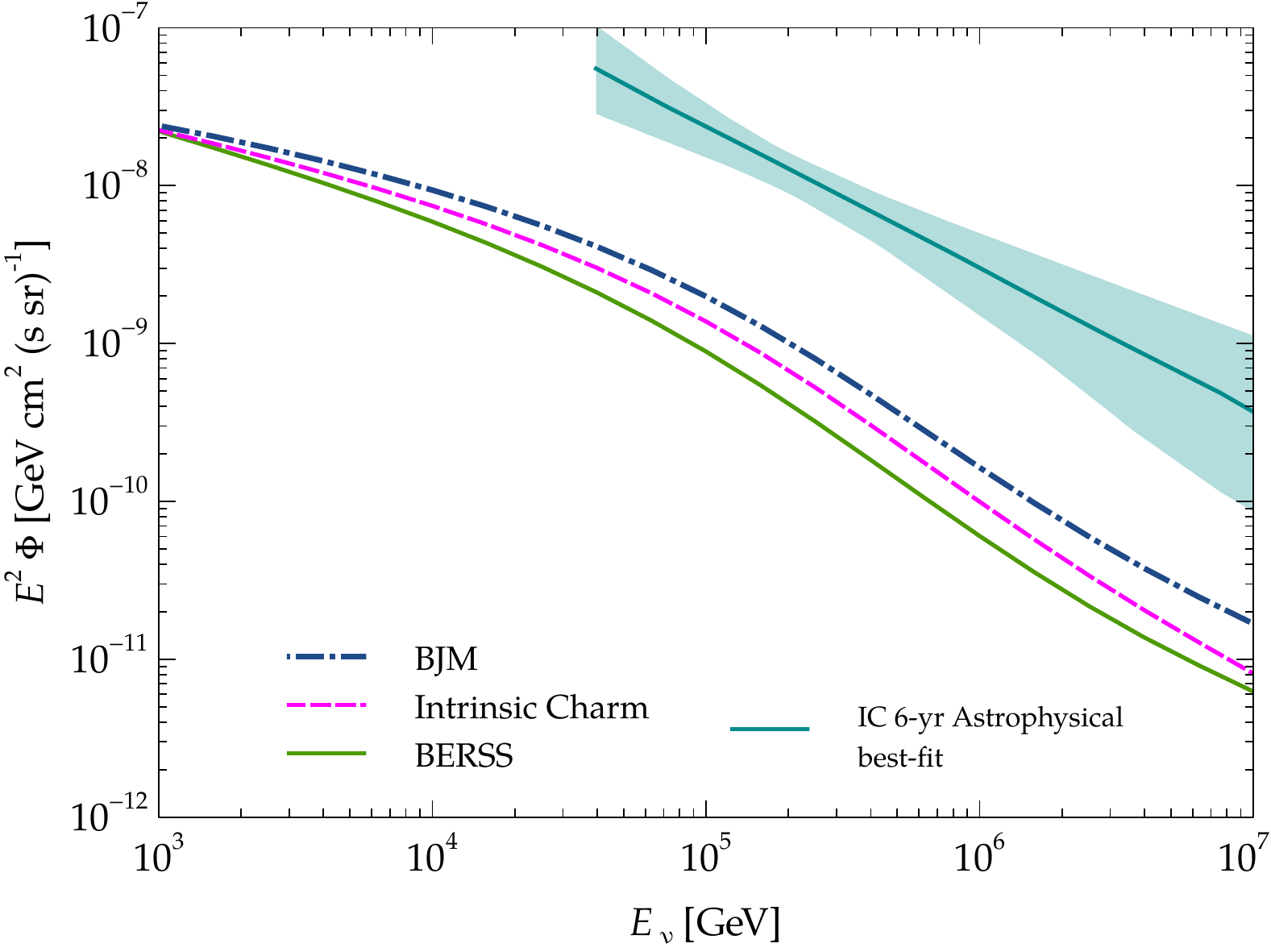}
  \caption{\label{fig:flux-ic}The results for prompt fluxes when including forward charm
           production from different formulations.}
\end{figure}

\paragraph*{}
Using the BJM cross-sections with the best-fit parameters above, we can determine the
corresponding neutrino flux expected at Earth as a consequence of interactions of cosmic rays
with atmospheric nuclei assumed to contain $ A $ nucleons.
As an estimate of the incident cosmic-ray proton flux, we use the models of Gaisser \cite{Gaisser:2011cc},
specifically his proton-rich estimates designated as $ H3p $.
The computation of the corresponding neutrino flux follows the semi-analytical procedure
outlined in standard literature (see \eg, \cite{Gaisser:1990vg}, \cite{Lipari:1993hd}).
Briefly, working in the exponential atmosphere approximation,
where the column density as a function of height is given by $ \rho(h) = \rho_0 \exp(h/h_0) $
with $ \rho_0 = 2.03 \times 10^{-3} \text{ g/cm}^3 $ and $ h_0 = 6.4 $ km,
the low- and high-energy lepton fluxes may be expressed in simple
semi-analytic forms in terms of spectrum-weighted $ Z $-moments.
These $ Z $-moments relate to the conversion of the incoming proton content
in the cosmic-ray flux to the heavy-meson flux produced therefrom ($ Z_{ph} $), and
then from the latter to the final leptonic flux reaching the detector ($ Z_{h\ell} $).
Additional moments $ Z_{pp} \text{ and } Z_{hh} $ describe respectively the energy losses of the protons in collisions
with air nuclei not leading to meson production, and the energy
losses of the meson before their decay resulting in leptons .
The full procedure is described in \cite{Bhattacharya:2015jpa}, and, for brevity, we refer
the reader to the discussion in Sec.\ 3 thereof rather than repeat it here.
Using this machinery, but with additional contributions to $ D^{\pm,0} $ production from
non-perturbative diffractive processes in the forward $ x_F $ region,
we have computed the total prompt neutrino flux in the intrinsic-charm and  scenarios.
These are shown in Fig.~\ref{fig:flux-ic}.

Note that the NLO amplitude does not interfere with the BJM amplitude
given that they fragment differently and, hence, the two final states are different.
In the intrinsic charm case, the kinematics of the charm quarks prevents significant
interference.
As far as BJM and intrinsic charm go, the cross-sections could in principle be added;
however, the fits with intrinsic charm structure functions do not include the BJM
mechanism and these two processes are, therefore, not independent.
The inclusion of the BJM mechanism in the fits would lower the contribution of intrinsic charm at high $x_F$.
 
\section{\label{sec:events}Event rates}
IceCube has been steadily accumulating events over the last seven years.
Its latest results \cite{Aartsen:2015zva} see 56 high-energy starting events within the energy range of 10 TeV--2.1 PeV.
By looking at the angular distribution of the incident background neutrinos, IceCube is
capable of distinguishing between those from the prompt flux, which shows a largely flat distribution,
and those from the conventional flux which is dominated by the vertical flux and suppressed
toward the horizons.
With present data, IceCube sees no evidence of the prompt flux yet, and accordingly sets a 90\% confidence level
upper bound at about 0.52 times the best-fit from \cite{Enberg:2008te}.
The total prompt flux, even when including that from the BJM recombination and intrinsic charm,
is consistent with this limit.
Future analyses, involving more data and possibly improved by incorporating ``self-veto'' methods to distinguish between signal
and background neutrinos \cite{Arguelles:2018awr}, are expected to improve this limit. 

\begin{figure}[htb]
  \centering
  \includegraphics[width=0.8\textwidth]{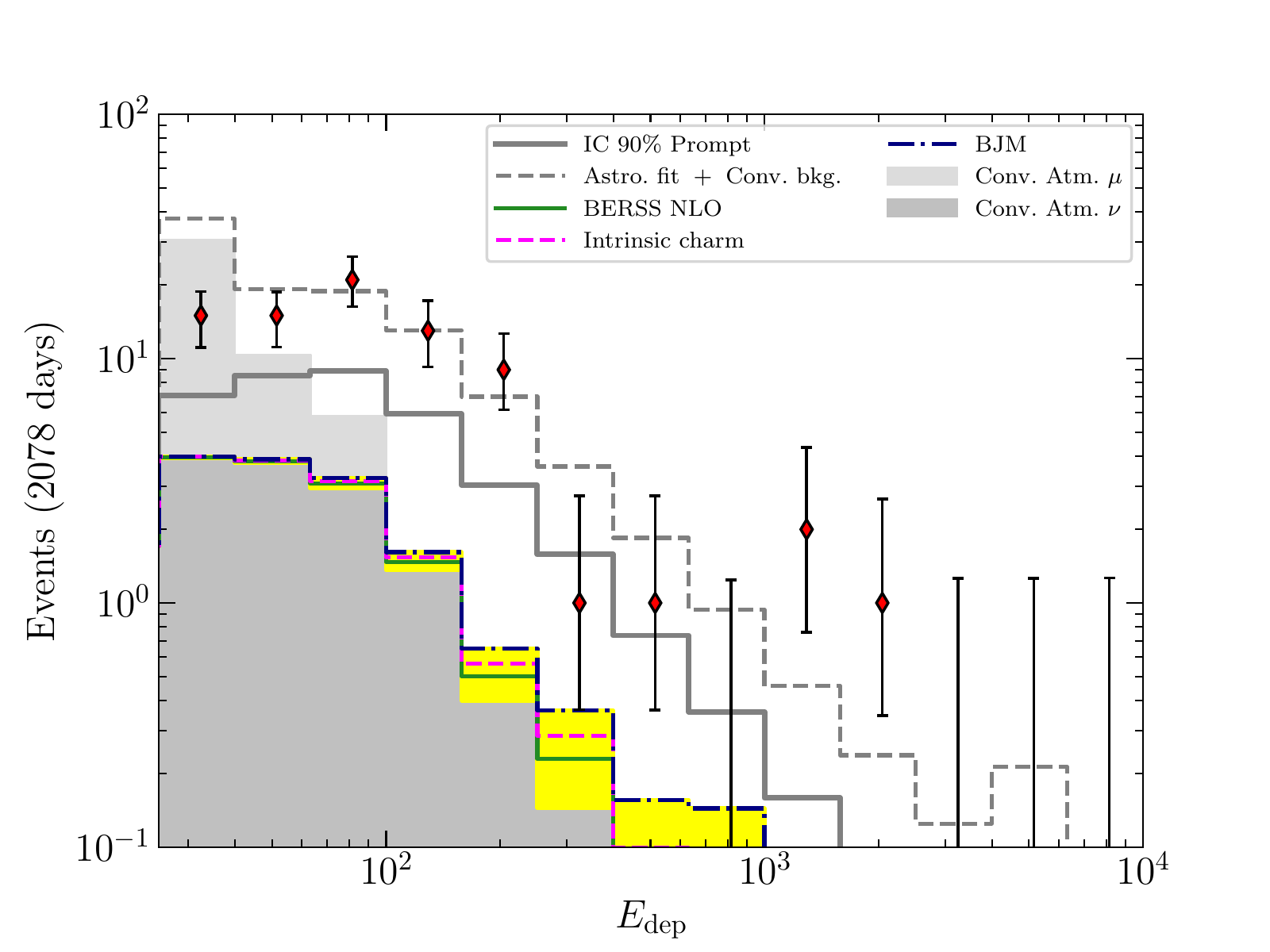}
  \caption{\label{fig:prompt-events} Event rates at IceCube showing the total atmospheric background (prompt +
           conventional) when different intrinsic charm model contributions are taken into account.}
\end{figure}

The overall modification to the prompt atmospheric neutrino background in terms of IceCube 6-year
event rates plotted against the energy ($ E_\text{dep} $) deposited in the detector,
courtesy interactions of the lepton with detector nuclei, is shown in \figref{prompt-events}.
While we have not re-evaluated the modification to the statistical significance of the non-atmospheric
signal in light of the modified background, it is clear from looking at the figure that the change will be
negligible.

\section{Conclusions}
We have evaluated the upper limit to the contribution to prompt neutrino background from
diffractive forward-rapidity cross sections by surveying existing models in the literature.
As the rapidities where such contributions can be significant are limited in range,
\textit{viz.} at very high $ x_F $, their contribution to the overall
prompt-neutrino flux is limited to the very high energies at IceCube $ E \gtrsim 200 $ TeV.
As such, the background, even when accounting for novel diffractive production mechanisms, is a rather minor player
in comparison to the flux of non-terrestrial neutrinos.
We have evaluated the upper limit to this component, maintaining consistency with
constraints at low and middle $ x_F $ from accelerator experiments, and have estimated the expected event rates.

It is evident, from analyses at IceCube and ultra-high-energy  cosmic-ray observatories, that
an understanding of the origin of  the extragalactic particles seen at energies of hundreds of TeV
and higher requires quantitatively enhanced data,
and more precision estimates of the various ingredients involved in their theoretical modelling.
The background is an important part of this understanding.

We have shown here that there are large uncertainties in the QCD modelling of the prompt signal.
However, in all considered cases, the contribution of novel mechanisms does not contribute
significantly to the prompt neutrino signal.

\acknowledgments{We thank Gr{\'e}gory Soyez and Jean-Philippe Lansberg for useful discussions.
We also thank the referee for helping us improve the discussions around our results.
This work was supported by the Fonds de la Recherche Scientifique-FNRS, Belgium, under grant number 4.4501.15.
AB is supported by the Fonds de la Recherche Scientifique de Belgique (F.R.S.-FNRS)
as ``Chargé de Recherches'' under grant number 1.B040.19.
AB is also thankful to the computational resource provided by Consortium des Équipements de Calcul Intensif (C{\'E}CI), funded by the F.R.S.-FNRS under grant number 2.5020.11 where a part of the computation was carried out.}

\bibliographystyle{JHEP}
\bibliography{incharmrefs}

\end{document}